\documentclass[runningheads]{svmult}
\usepackage{graphicx}  
\usepackage{multicol}  
\usepackage{physprbb}  

\usepackage{units} 
\usepackage{booktabs} 

\usepackage{amsmath,amssymb,latexsym}
\begin{document}
\pagestyle{headings}  
\mainmatter           
\title*{Online traffic state estimation based on floating car data}
\titlerunning{Online traffic state estimation}
%
\author{Arne Kesting\inst{1} \and Martin Treiber\inst{1}}
\authorrunning{A. Kesting and M. Treiber}   
\institute{Technische Universit\"at Dresden, Institute for Transport \& Economics,\\
  W\"urzburger Str. 35, 01062 Dresden, Germany}

\maketitle              

\begin{abstract}
Besides the traditional data collection by stationary detectors, recent advances in wireless and
sensor technologies have promoted new potentials for a vehicle-based
data collection and local dissemination of information. By means of microscopic traffic simulations we study the problem of online estimation of the current traffic situation based
on floating car data. Our focus is on the estimation on the up- and downstream jam fronts
determining the extension of traffic congestion. We study the impact
of delayed information transmission by short-range communication via
wireless LAN in contrast to instantaneous information transmission to
the roadside units by means of mobile radio. The delayed
information transmission leads to systematic estimation errors which
cannot be compensated for by a higher percentage of probe vehicles. Additional flow measurements from stationary detectors allow for a model-based prediction which is effective for much lower floating car percentages than \unit[1]{\%}.
\end{abstract}

\section{\label{sec:intro}Introduction}
%
Detailed and reliable online traffic state estimation is a
prerequisite for advanced traffic information systems and the next
generation of driver assistance systems~\cite{Driel-ADAS-impact-2007,Arne-ACC-TRC}. 
Recent advances in wireless and sensor technologies have therefore promoted new
potentials for a vehicle-infrastructure integration (VII) system which allows for a wireless communication between roadside sensors and vehicles equipped with communication interfaces~\cite{USDOT-ITS,USDOT-IntelliDrive}. Data gathered from equipped vehicles (`probe vehicles' or `floating cars') can be used to assess and predict traffic conditions and, in turn, information about incidents, travel times or congestion can be communicated from roadside units to the vehicles~\cite{Ma-C2I-2009,Tanikella-VII-2007,Nanthawichit-ProbeVeh-03,schaefer2002traffic}. 

In this paper, we consider the following setup for an online traffic-surveil\-lance application based on floating car data: Probe vehicles collect their positions and speeds over time and communicate this information to roadside units either periodically by instantaneous forwarding by mobile phone communication, or, alternatively, by short-range communication when passing the location of a roadside unit. In any case, the roadside units are connected to each other and collect the data periodically for an online traffic state estimation. More specifically, we consider the prediction of the upstream jam front which, for example, can be used to warn the driver when congestion is ahead. For this purpose, we study the data collection, the communication to roadside units, and the traffic state estimation `in-the-loop' by means of microscopic traffic simulations as a function of the percentage of equipped vehicles. This approach enables us to compare the estimates of both communication modes to the -- known and reproducible -- traffic situation. Since the expected equipment levels in the first deployment phase will be small, we consider small percentages of probe vehicles below \unit[3]{\%} only.

The paper is structured as follows: In Sec.~\ref{sec:simSetup}, the microscopic simulation set-up is described. In Sec.~\ref{sec:Estimation}, the algorithm for the estimation of jam fronts is presented and a measure for evaluating the estimation quality is defined. In Sec.~\ref{sec:calcVg} an alternative model-based approach is presented which uses additionally flow measurements from stationary detector. We close with a discussion in Sec.~\ref{sec:diss}.

\section{\label{sec:simSetup}Microscopic Simulation Setup}
%
\subsection{\label{sec:scen}Reference Scenario with Empirical Boundary Conditions}
\begin{figure}
\centering
\begin{tabular}{cc}
\includegraphics[width=0.5\textwidth]{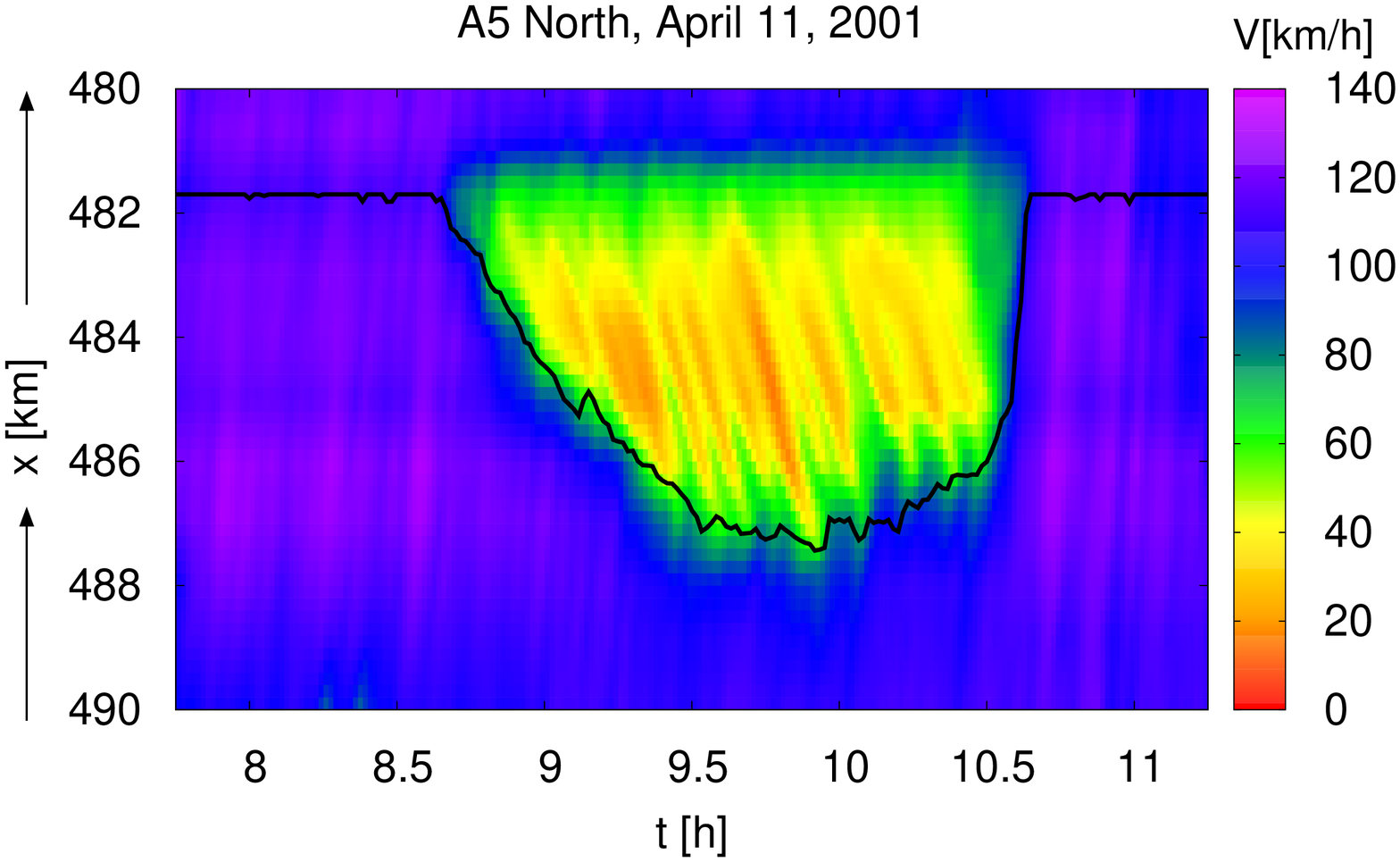} &
\includegraphics[width=0.5\textwidth]{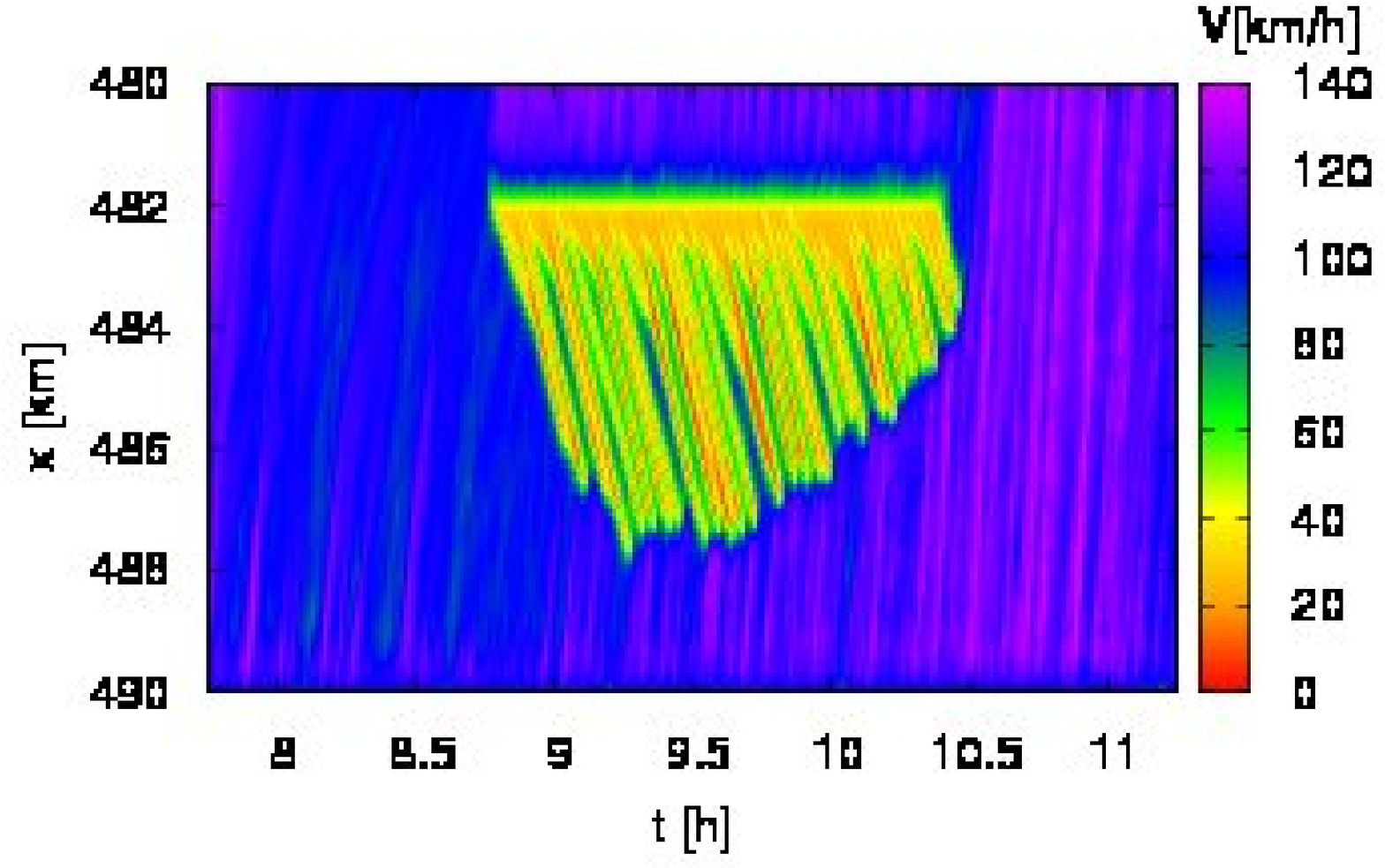} 
\end{tabular}

 \caption{\label{fig:trafficSituation}Empirical traffic state from the German freeway A5 reconstructed from stationary loop detector data with an dedicated interpolation method~\cite{ASM-CACAIE} (left) and spatiotemporal traffic dynamics in a calibrated multi-lane simulation (right). }

\end{figure}

\begin{figure}
\centering
\begin{tabular}{cc}
\includegraphics[width=0.5\textwidth]{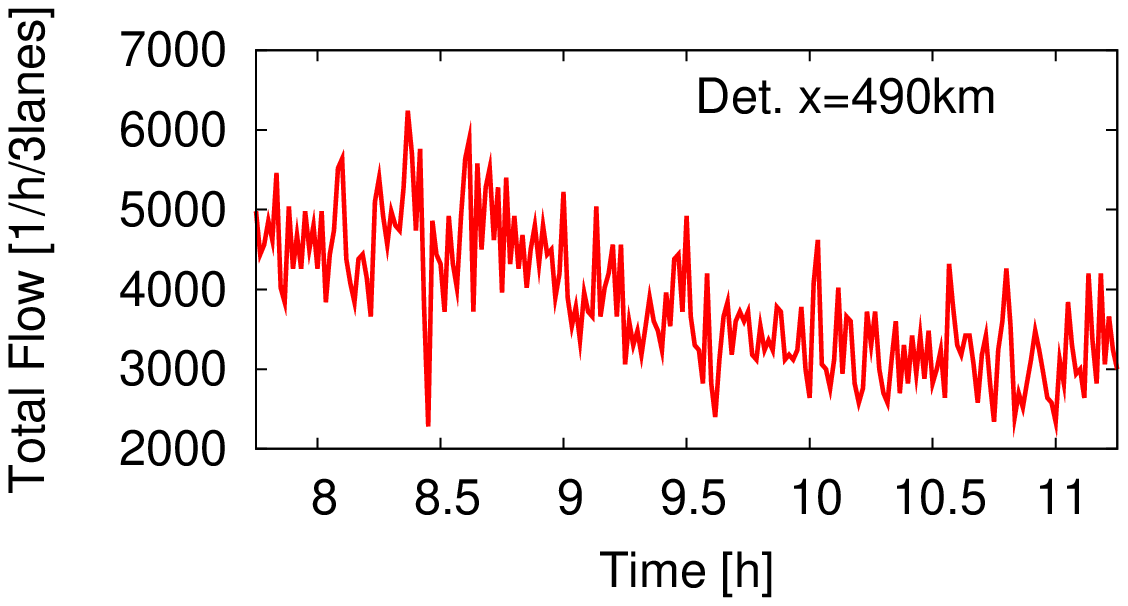} &
\includegraphics[width=0.5\textwidth]{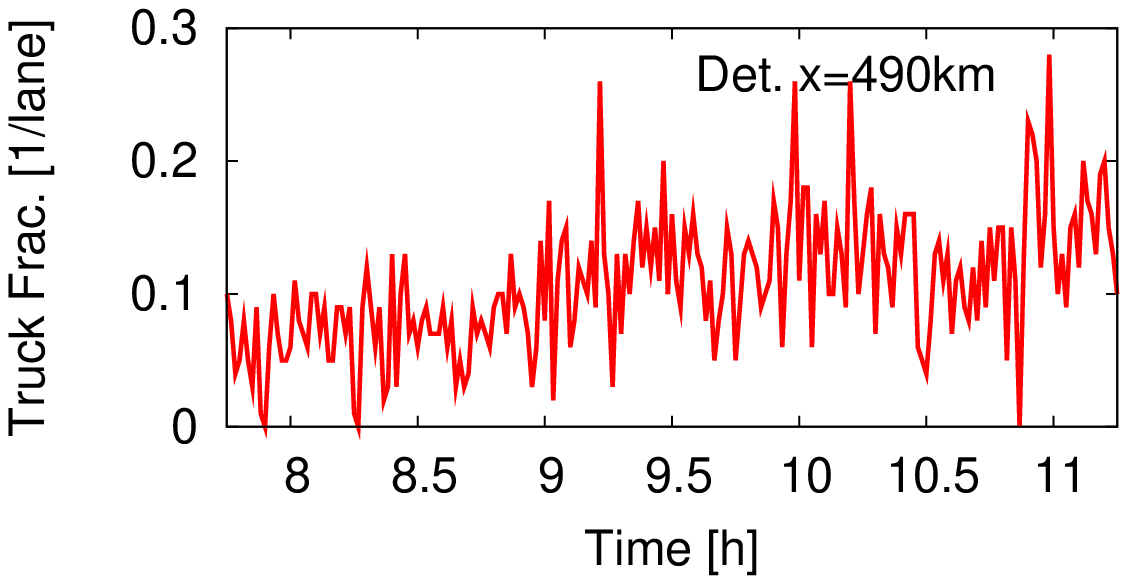} 
\end{tabular}

 \caption{\label{fig:boundaryCond}Timeseries of traffic flow and truck fraction from the German freeway A5 used as upstream boundary conditions in the simulation.}

\end{figure}

For the following case study of online estimation of jam fronts one needs a suited and realistic traffic situation serving as reference scenario. For this purpose we start with the empirical traffic jam observed on the German freeway A5 shown in Fig.~\ref{fig:trafficSituation}(left). In the simulation environment, we consider a road-section with three lanes and a flow-conserving bottleneck~\cite{Opus}. The empirical traffic flows and truck percentages  from one detector location (shown in Fig.~\ref{fig:boundaryCond}) serve as upstream boundary conditions assuring realistic traffic volumes and degrees of traffic heterogeneity. 
The simulator uses the \textit{Intelligent Driver Model}~\cite{Opus} as a simple, yet
realistic, car-following model, and the general-purpose lane-changing
algorithm MOBIL~\cite{MOBIL-TRR07}. The parameters of the car-following model (see Table~\ref{tab:IDM}) and the bottleneck strength (modeled by a local increase of the time gap parameter $T$ by \unit[40]{\%} in a given time interval) have been adapted in order to reproduce the empirically observed spatiotemporal congestion pattern on a semi-quantitative level. Figure~\ref{fig:trafficSituation}(right) shows the simulation result.

\begin{table}

\newcommand{\entry}[3]{
  \parbox{50mm}{\vspace*{1mm}#1\vspace*{1mm}}
  &\parbox{22mm}{\vspace*{1mm}\centering{#2}\vspace*{1mm}}
  &\parbox{22mm}{\vspace*{1mm}\centering{#3}\vspace*{1mm}} }

\centering
\begin{tabular}{p{45mm}  p{18mm} p{14mm}}
\toprule
Parameter  &  Car &  Truck\\ 
\midrule
Desired speed $v_0$  &  \unit[150]{km/h}   &  \unit[90]{km/h} \\
Desired time gap $T$       &  \unit[1.15]{s}   &   \unit[2.1]{s}\\
Jam distance $s_0$      &  \unit[2.5]{m}    &   \unit[4.0]{m} \\
Gap parameter $s_1$     &  \unit[3.0]{m}    &   \unit[3.0]{m} \\
Max. acceleration $a$     &  \unit[1.0]{m/s$^2$}   &   \unit[0.8]{m/s$^2$} \\
Comfortable deceleration $b$   &  \unit[2.0]{m/s$^2$}   &   \unit[2.0]{m/s$^2$} \\
\bottomrule
\end{tabular}

 \caption{\label{tab:IDM}Model parameters of the Intelligent Driver Model (IDM)~\cite{Opus} used in the reference scenario shown in Fig.~\protect\ref{fig:trafficSituation}. The vehicle length is~\unit[8]{m} for cars and~\unit[15]{m} for trucks.}

\end{table}

\subsection{\label{sec:simFCD}Floating Car Samples}

\begin{figure}
\centering
\begin{tabular}{cc}
\includegraphics[width=0.5\textwidth]{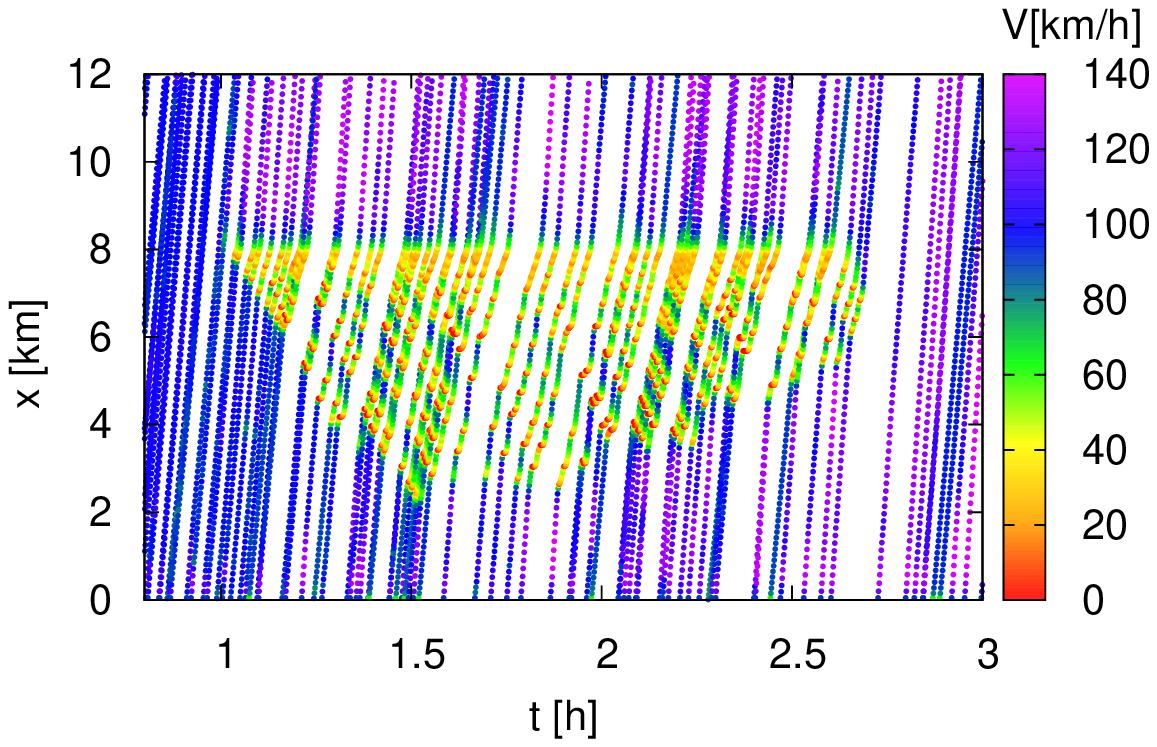} &
\includegraphics[width=0.5\textwidth]{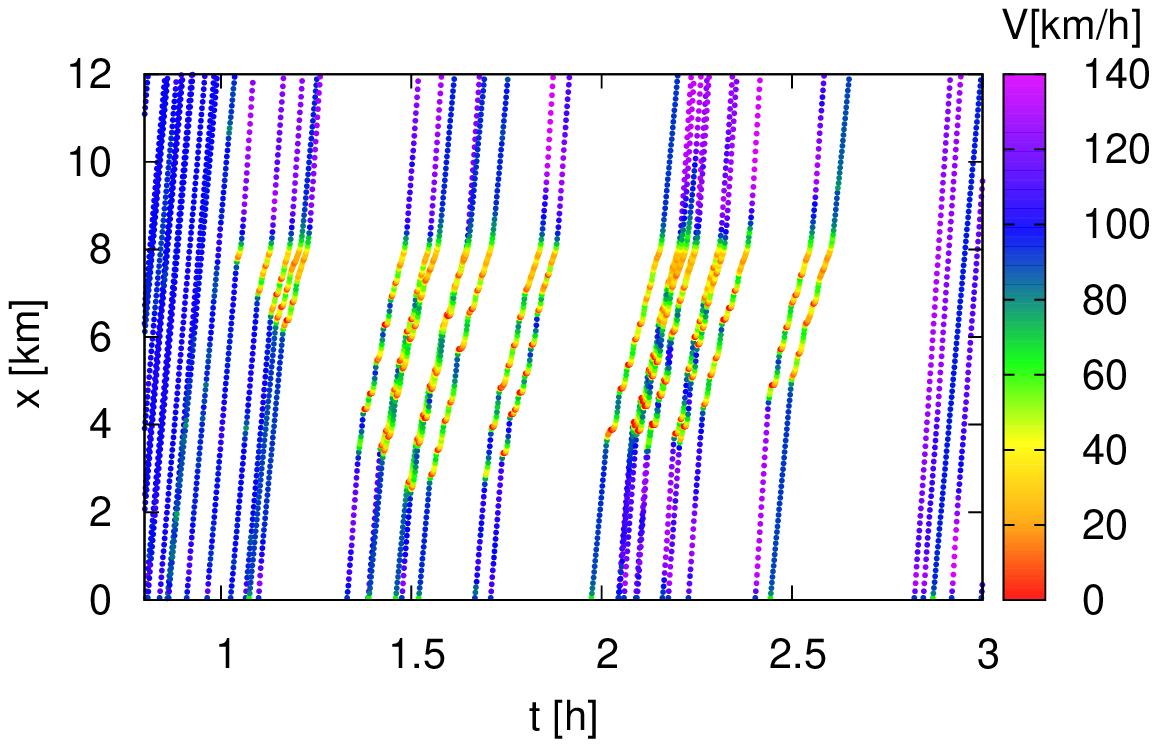} 
\end{tabular}

 \caption{\label{fig:fcdSamples}Random samples of floating cars for equipment levels of \unit[1]{\%} (left) and \unit[0.5]{\%} (right).}

\end{figure}

In the microscopic simulation, a given percentage of vehicles is randomly selected to generate floating-car data. These vehicles record their positions and speeds with a period of~\unit[5]{s}. The random seed has been fixed in order to assure identical samples of floating cars (FC). In total, $13\,394$~vehicles have been simulated in the three-lane scenario over \unit[3.5]{h}. 137~vehicles (63) have been selected for a given FC percentage of \unit[1]{\%} (\unit[0.5]{\%}), i.e., the sample realized by the seed effectively contain \unit[1.02]{\%} and \unit[0.47]{\%}, respectively. This corresponds to an \emph{average} time gap of 
\unit[1.5]{min} (\unit[3.3]{min}). Figure~\ref{fig:fcdSamples} shows the trajectories of these FC. In contrast to stationary detector data which are typically provided once per minute, time gaps between floating car are distributed randomly. For a constant traffic flow, these gaps can be described approximately by an exponential distribution~\cite{thiemann-IVC-PRE08}.

\subsection{\label{sec:commModes}Modes of Vehicle-to-Infrastructure Communication}

For transmitting the floating car data to the roadside units (RSU) we simulate the following operation modes which can be considered as the limiting cases of delay times in an idealized communication:
\begin{itemize}
\item
\textbf{Local transmission from the probe vehicle to the RSU:} All information is broadcasted when passing the cross-section. This corresponds to a short-range communication, e.g. by wireless communication interfaces with a minimal broadcast range (WLAN).
\item
\textbf{Instantaneous transmission:} After collection, data is directly transmitted to the RSU. This corresponds to a nonlocal communication, e.g. by using the cellular phone network (GSM).
\end{itemize}

In the following section, we study the impact
of delayed information transmission by short-range communication in contrast to instantaneous information transmission. Obviously, the time delay in the former case depends on the number of RSU and  their positions. Moreover, the delay will be determined by the local traffic conditions. For example, a severe traffic jam will have a significant impact on the up-to-dateness of incoming FCD.

\section{\label{sec:Estimation}Online Estimation of Jam Fronts}

After having defined the simulation setup, we now turn to the problem of using the {\it currently} available information at the RSU for predicting the positions of the jam fronts in an online application. Reliable and up-to-date information about these `shock fronts' can then be used for  applications in safety, driver information or traffic-adaptive cruise control systems~\cite{Arne-ACC-TRC}.
\subsection{\label{sec:weightedRegr}Extrapolation of Jam Fronts by Weighted Linear Regression}
\begin{figure}
\centering
\begin{tabular}{cc}
\includegraphics[width=0.5\textwidth]{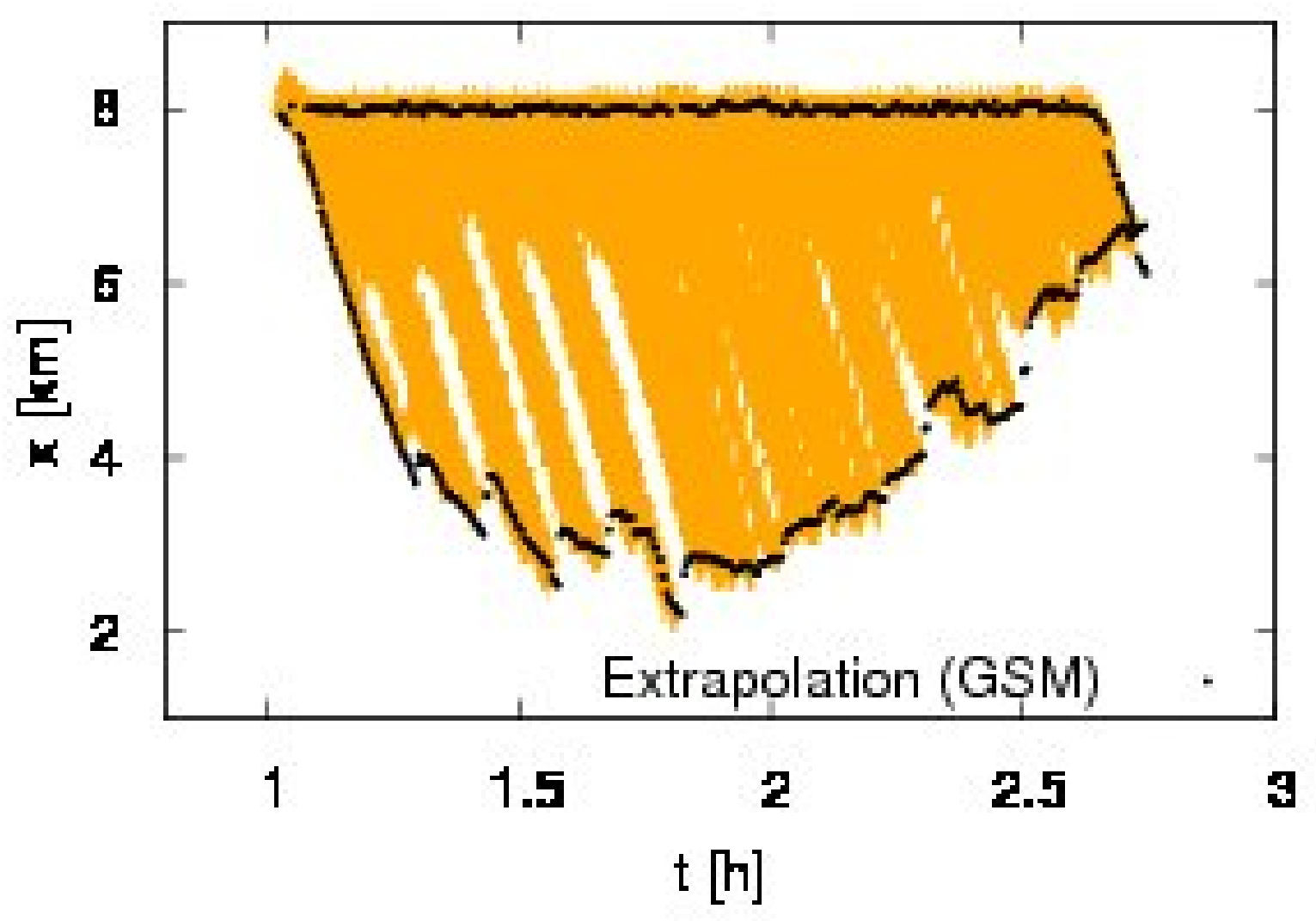} &
\includegraphics[width=0.5\textwidth]{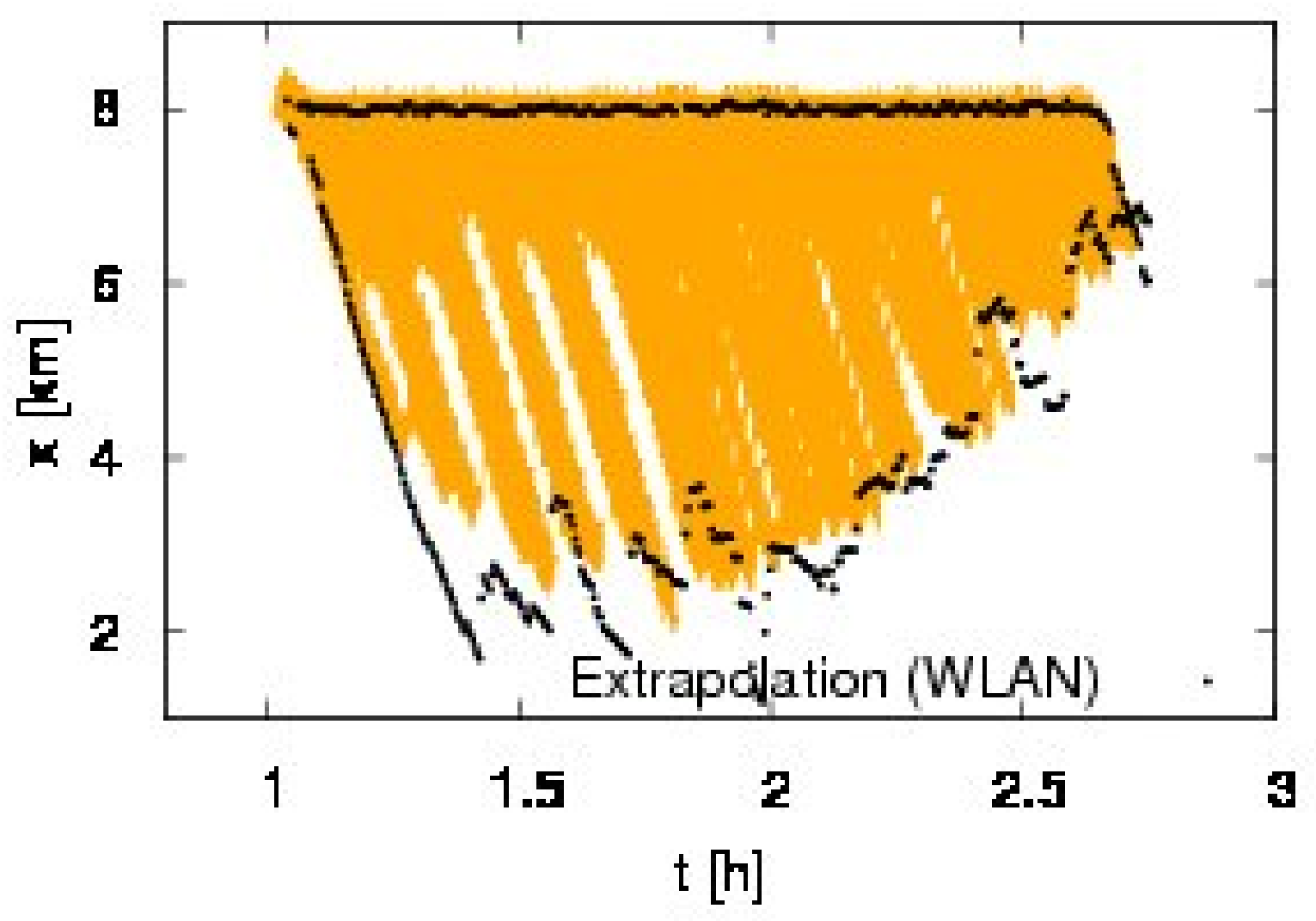} 
\end{tabular}

 \caption{\label{fig:extraExamples}Examples of extrapolated up- and downstream jam fronts with weighted regression using $\lambda=\unit[0.33]{min}^{-1}$. The `ground truth' traffic dynamics is shown for comparison. The delayed transmission by short-range wireless communication results in a less precise estimation of the upstream jam front (right) than in the case of instantaneous transmission (left). }

\end{figure}

When a floating car passes through the traffic jam, the first drop in its speed measurements below a certain threshold, e.g. \unit[50]{km/h}, defines one `spatiotemporal' point $(t_i, x_i)$ of the upstream jam front. 
After having gathered a sufficient number $n$ of points (the last point $n$ at time $t_n$), the position of the jam front at times $t>t_n$ can be predicted by a linear extrapolation which is based on a weighted linear regression. In order to favor newer data points over older ones, we use for the weights an exponential 
\begin{equation}\label{eq:leverage_function}
K(t-t_i) = e^{- \lambda (t-t_i)}.
\end{equation}
In the simulation, this `online' extrapolation is carried out for both communication modes. Examples of the estimated jam fronts are shown in Fig.~\ref{fig:extraExamples}.

\subsection{\label{sec:analysis}Evaluation of the Online Estimation Results}

The prediction of the downstream jam front is relatively easy since this front is either stationary (for activated bottlenecks) or propagates with a fixed velocity of \unit[-15]{km/h} against the direction of travel (cf. the traffic dynamics depicted in Fig.~\ref{fig:trafficSituation}). Thus, we focus on the estimation of the upstream jam front $x^\text{up}$ which does not have a characteristic propagation velocity but depends on the prevailing in- and outflows. As error quantity, we consider the deviation between the estimation $x_\text{est}^\text{up}$ derived from Eq.~\eqref{eq:leverage_function} and the simulation $x_\text{sim}^\text{up}$ (defined by a speed threshold of \unit[50]{km/h}): 
\begin{equation}\label{eq:deviation_fronts}
\Delta x(t) = x_\text{est}^\text{up}(t) - x_\text{sim}^\text{up}(t).
\end{equation}
From the distribution of the estimation errors $\Delta x(t_i), i=1\ldots n$, the mean deviation $\langle \Delta x \rangle$ of the estimated jam front is defined by averaging over all predicted positions. Furthermore, the variation $\sigma_x$ is a relevant  measure for its reliability.

\begin{figure}
\centering
\begin{tabular}{cc}
 \includegraphics[width=0.5\textwidth]{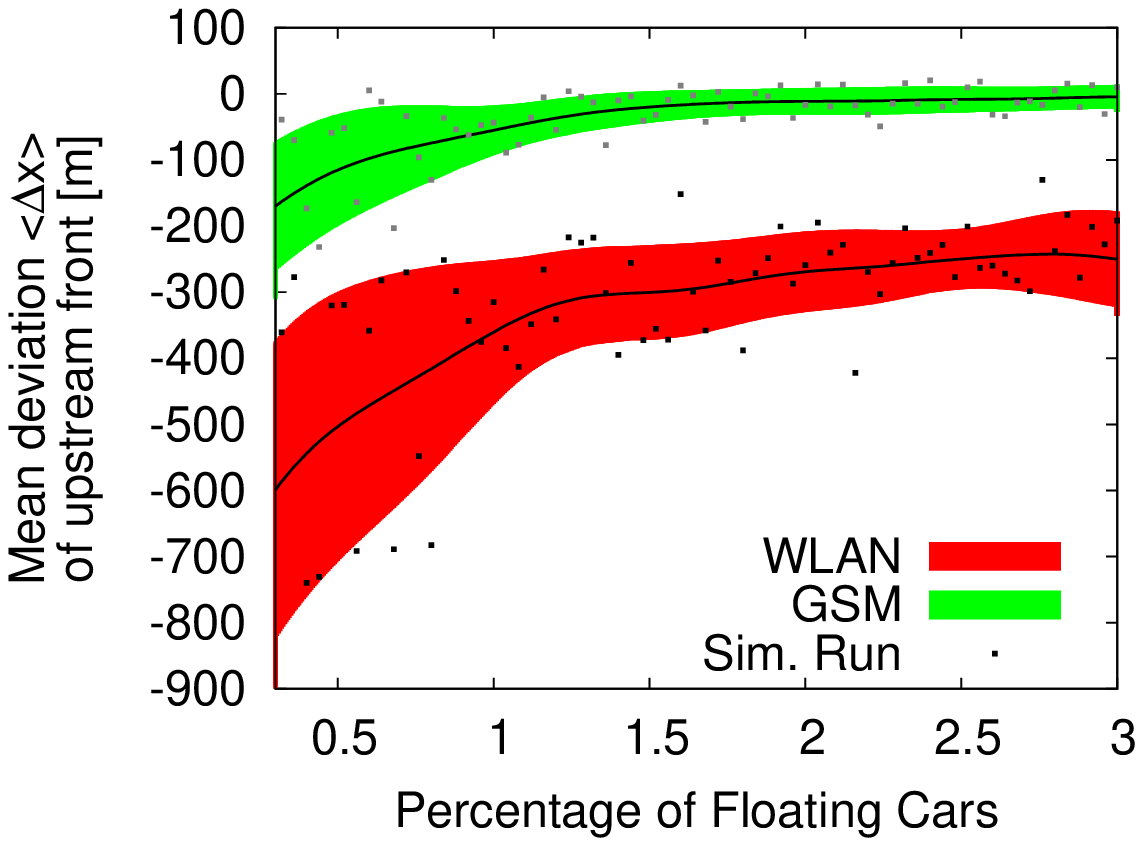} &
\includegraphics[width=0.5\textwidth]{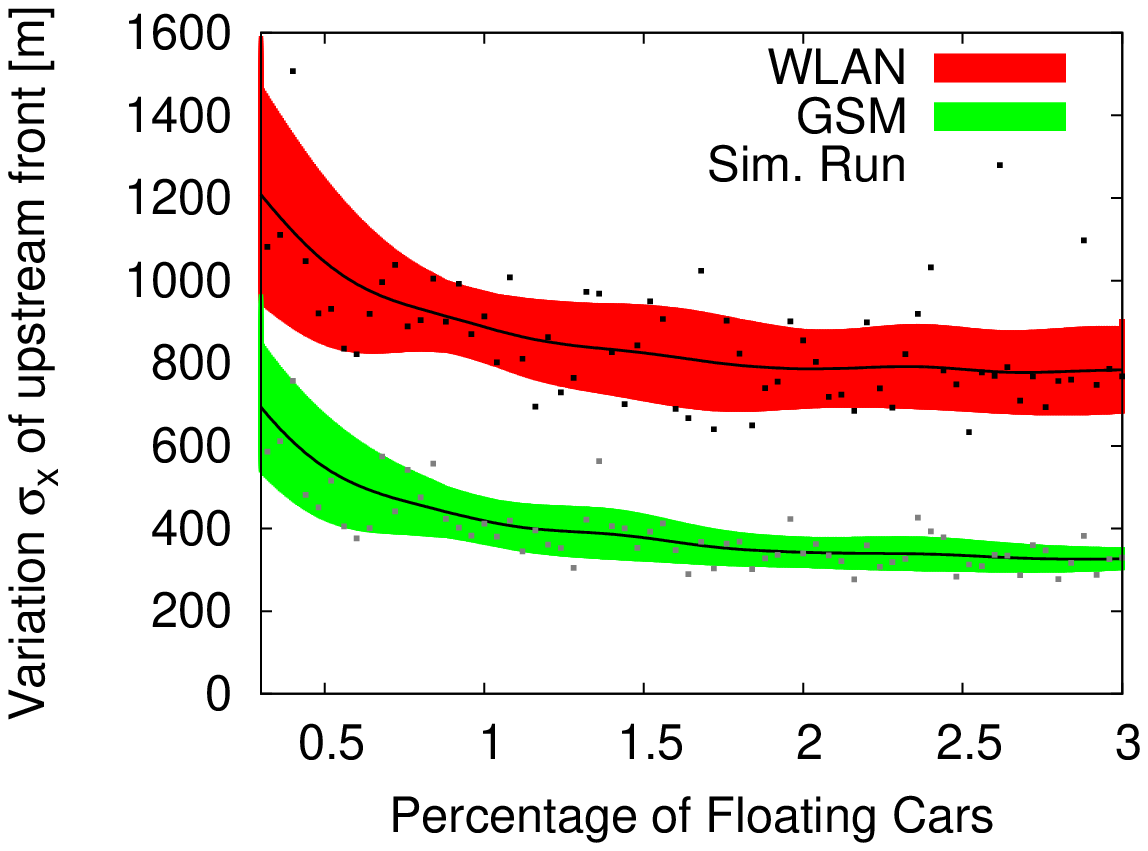} 
\end{tabular}

 \caption{\label{fig:resultsFC}Mean error and variation of estimated upstream jam front position. The percentage of equipped vehicles is systematically varied.}

\end{figure}

Figure~\ref{fig:resultsFC} shows the resulting mean deviation $\langle \Delta x \rangle$ and the variation $\sigma_x$ of the upstream jam front as a function of percentage of equipped vehicles for the instantaneous (GSM) and the delayed (WLAN) communication modes. The `bands' are calculated from single simulation runs (shown as single points) by a local linear regression method~\cite{kesting-acc-roysoc}. Both modes show the same cross-over behavior: Below a floating car fraction of about \unit[1.2]{\%}, the estimation errors increase with decreasing equipment level. Above \unit[1.2]{\%}, the prediction errors stay approximately constant indicating a diminishing benefit from additional (and thus redundant) information provided by floating cars for the chosen value of $\lambda$. For an equipment level of \unit[1.2]{\%}, the average time gap between two passing vehicles is about \unit[1]{min}.  In this case, the prediction error is about $(-36 \pm 400)\,\text{m}$ for the GSM mode and $(320 \pm 850)\,\text{m}$ for the WLAN mode. The large systematic errors in the prediction in the local information transmission is due to the large delays of up to \unit[10]{min} before equipped vehicles can broadcast the information to the RSU located downstream the traffic jam. For a local vehicle-infrastructure communication, this dominating time delay can only be compensated for by increasing the number of RSU along the road.  

\section{\label{sec:calcVg}Model-based Estimation using Additional Flow Data}

\begin{figure}
\centering
\begin{tabular}{cc}
\includegraphics[width=0.5\textwidth]{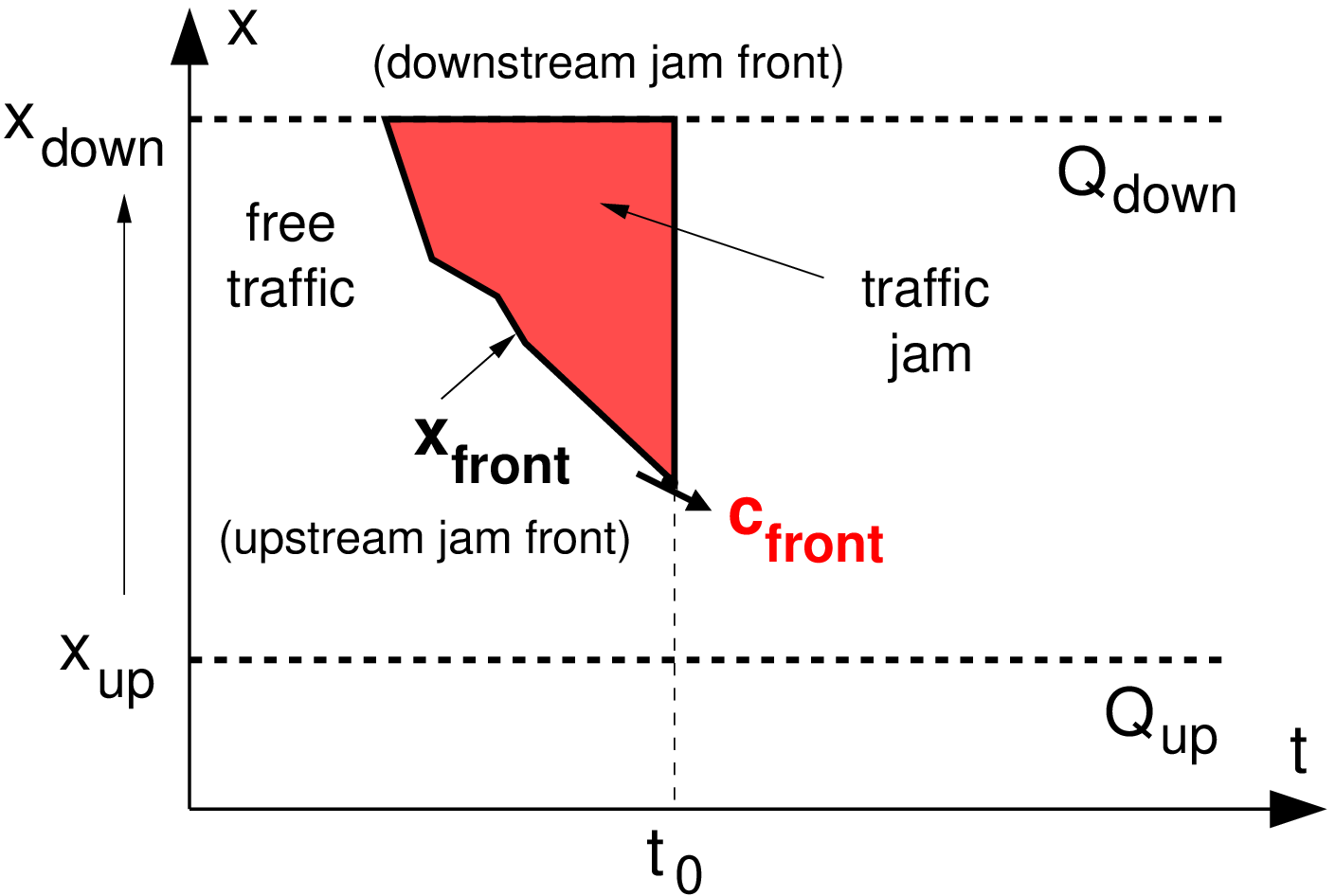} &
\includegraphics[width=0.5\textwidth]{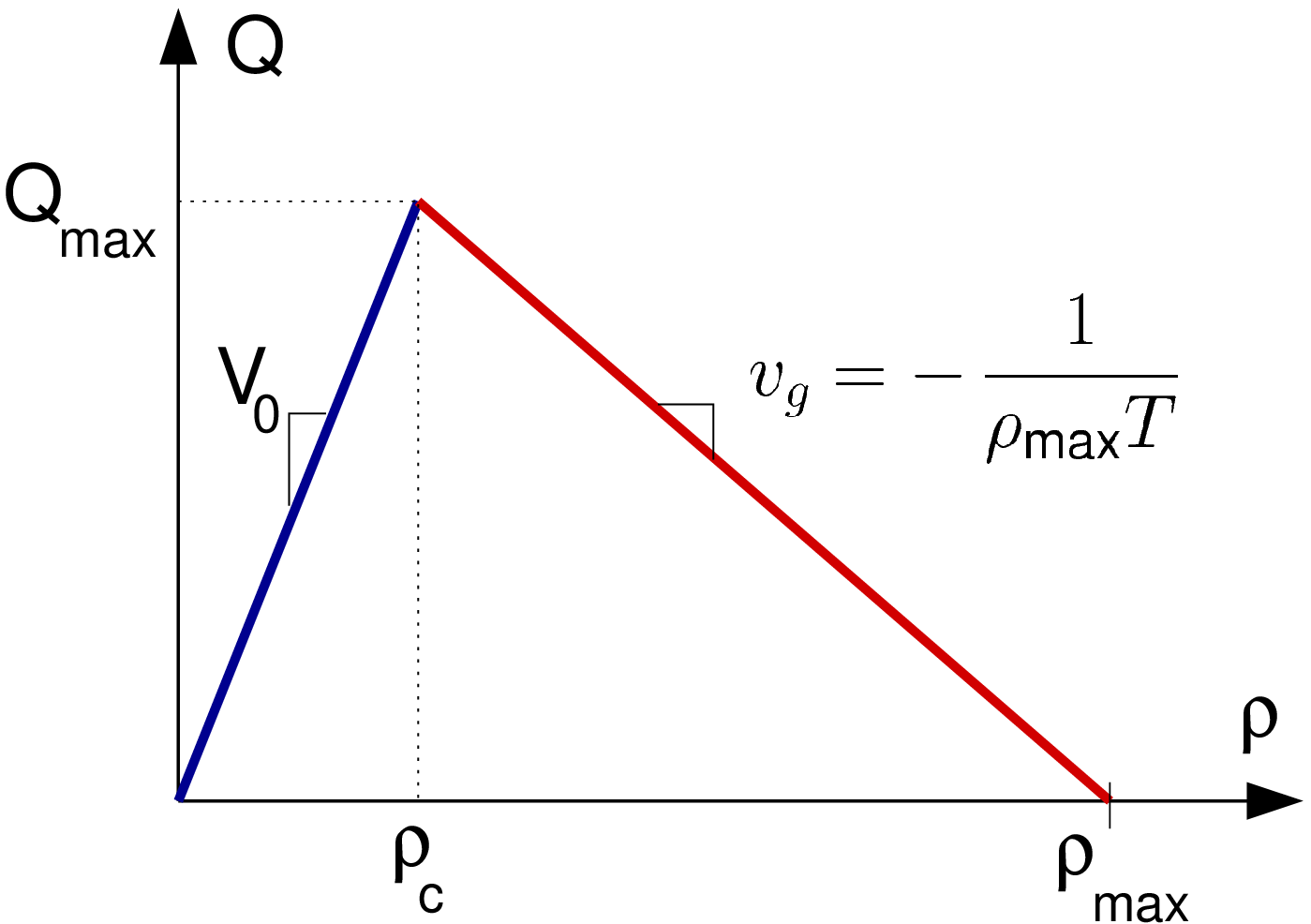} 
\end{tabular}

 \caption{\label{fig:calcVg}Left: Sketch of the considered detectors located up- and downstream of the traffic jam. Right: Assumed flow-density relationship ({\it fundamental diagram}) with two propagation speeds only.}

\end{figure}

The online detection of the upstream jam front solely from the speed information of floating vehicles is only possible with prediction errors particularly when considering a local vehicle-infrastructure communication by wireless interfaces to only one RSU located downstream a bottleneck. In order to improve the estimation quality, we additionally consider flow data from stationary detectors. Since the number of vehicles is a conserved quantity, model-based approaches using the continuity equation can be applied. A consequence of continuity is the well-known propagation equation for shock fronts  
\begin{equation}\label{eq:cfront}
c_\text{front} = \frac{\text{d}x_\text{front}}{\text{d}t} = \frac{Q_2 - Q_1}{\rho_\text{cong}(Q_2) - \rho_\text{free}(Q_1)}\,.
\end{equation}
Thus, the propagation speed of the upstream jam front can be calculated if the flows $Q_1$ and $Q_2$ of either side of the jam front are known (see left sketch in Fig.~\ref{fig:calcVg}), and if one additionally assumes a \emph{fundamental diagram}, i.e. a functional relation between traffic flow $Q$ and density $\rho$. As a simple model, we consider a `triangular' fundamental diagram (see right plot in Fig.~\ref{fig:calcVg}) with only two propagation velocities: (i) $V_0$ is the speed in free traffic conditions and corresponds to the mean travel speed of the vehicles (with a value of the order of \unit[100]{km/h}. (ii) In contrast, perturbations in congested traffic propagate against the direction of travel with a (remarkably constant) velocity of about $v_g=\unit[-15]{km/h}$. By using both velocities, the (flow) information measured at the detector positions $x_\text{up}$ and $x_\text{down}$ can be connected to the current upstream jam front position $x_\text{front}$:
\begin{eqnarray}
Q_1(t) &=  Q_\text{free}(x_\text{front}, t)  =& Q_\text{up} \left(t-\frac{x_\text{front} - x_\text{up} }{V_0} \right),\\
Q_2(t) & =  Q_\text{cong}(x_\text{front}, t) =& Q_\text{down} \left(t - \frac{x_\text{down}-x_\text{front}}{|v_g|} \right)\,.
\end{eqnarray}
Interestingly, both detector measurements are requested at {\it past} times which qualifies this approach for an online application in real-time. 

\begin{figure}
\centering
\begin{tabular}{cc}
\includegraphics[width=0.5\textwidth]{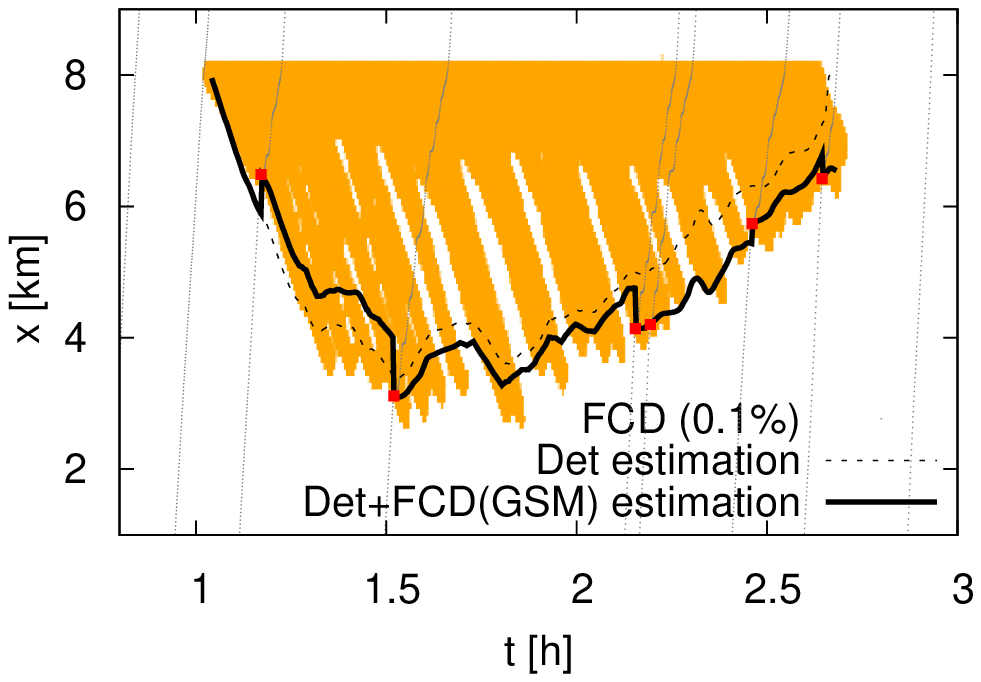} &
\includegraphics[width=0.5\textwidth]{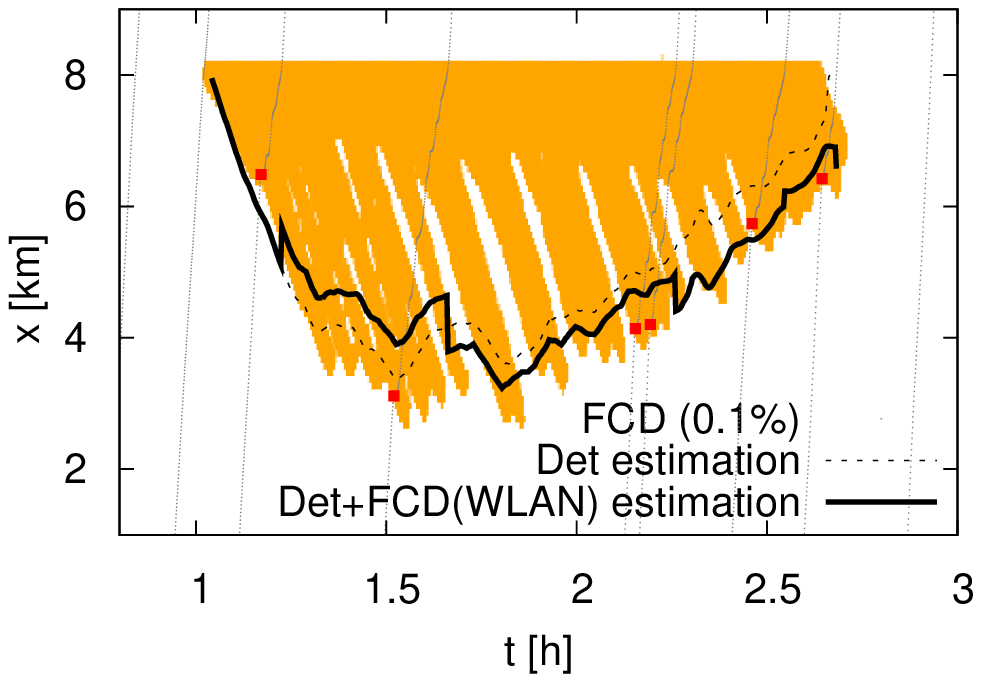}  \\[-8mm]
\includegraphics[width=0.5\textwidth]{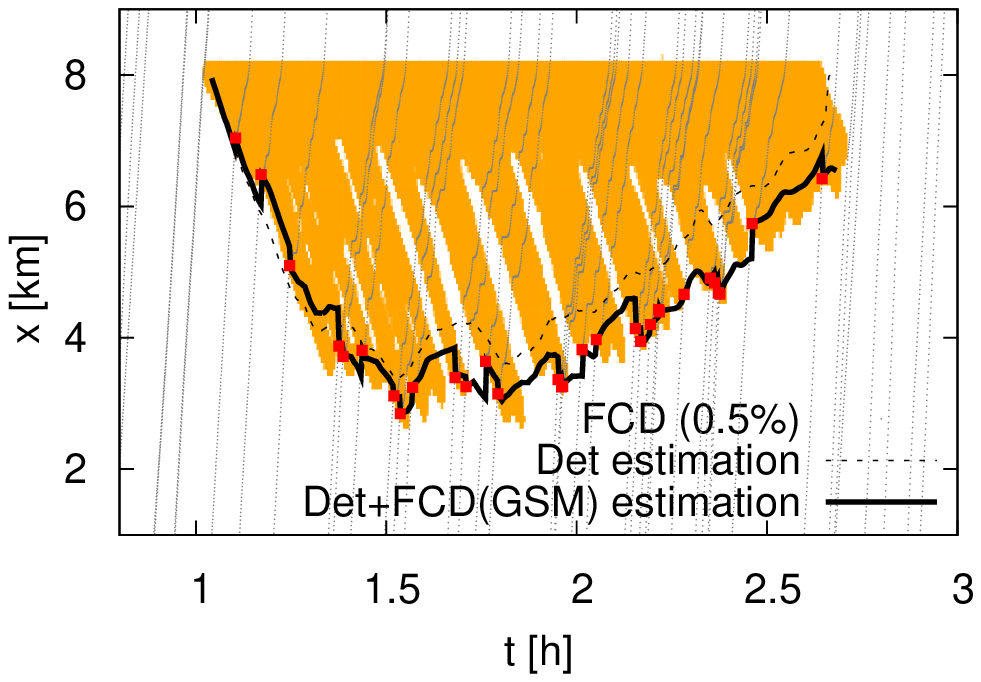} &
\includegraphics[width=0.5\textwidth]{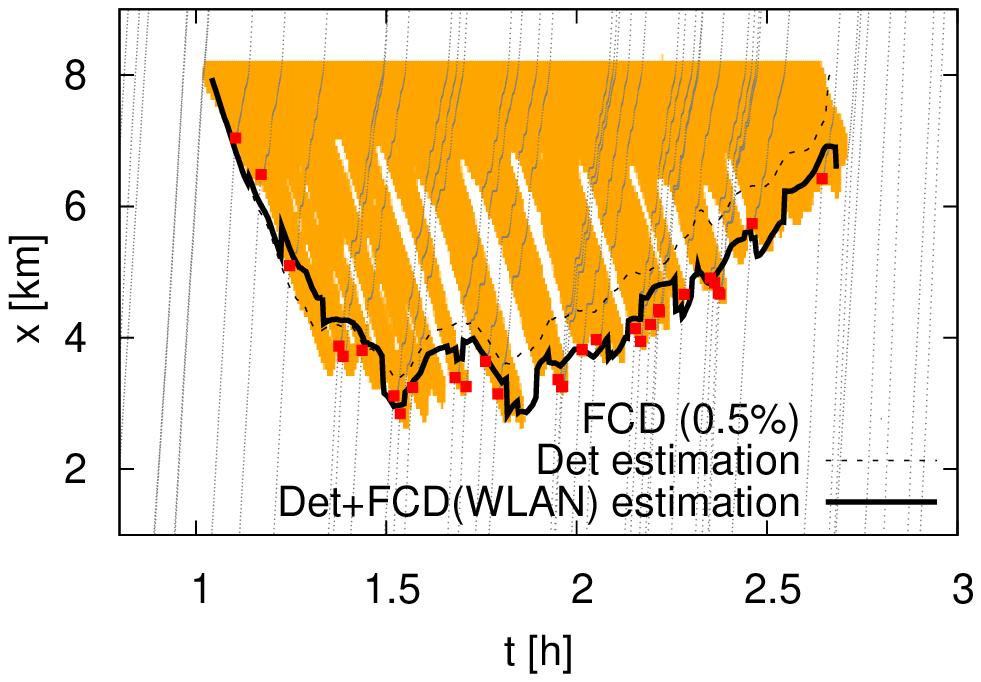} 
\end{tabular}

 \caption{\label{fig:exampleCombi}Estimation of upstream jam front using a combination of speed data from floating cars (\unit[0.1]{\%} and \unit[0.5]{\%}) and flow data (1-min aggregation interval) from stationary detectors located up- and downstream the traffic jam. }

\end{figure}

In Fig.~\ref{fig:trafficSituation}, the proposed prediction model has been applied to the empirical traffic data using the calibrated parameters $T=\unit[1.8]{s}$ (time gap), $\rho_\text{max}=\unit[80]{/km}$ (maximum density), and $V_0=\unit[120]{km/h}$ (desired free speed). Here, the starting point for the integration of Eq.~\eqref{eq:cfront} has been set manually. In combination with available FCD, this procedure can be automated because the first floating car that detects the traffic breakdown triggers the integration. Furthermore, the current position of the upstream jam front $x_\text{front}(t)$ detected by a floating car at (a past) time $t$ can be used for an automated correction of the prediction deviations which are summing up by errors in the jam front velocity $c_\text{front}(t)$. Furthermore, the deviations can be used as input to an online calibration of the model parameters. Whenever the jam front position has been reset at time $t$, the ODE~\eqref{eq:cfront} has to be integrated again up to the actual time $t' > t$.  Figure~\ref{fig:exampleCombi} shows the prediction results of the upstream jam fronts for FC equipment percentages of \unit[0.1]{\%} and \unit[0.5]{\%} using the uncalibrated parameters $T= \unit[2.0]{s}$, $\rho_\text{max} = \unit[100]{/km}$, and $V_0 =\unit[100]{km/h}$. For the purpose of illustration, the jam front prediction without automated FC correction is plotted as thin dashed lines.

\begin{figure}
\centering
\begin{tabular}{cc}
\includegraphics[width=0.5\textwidth]{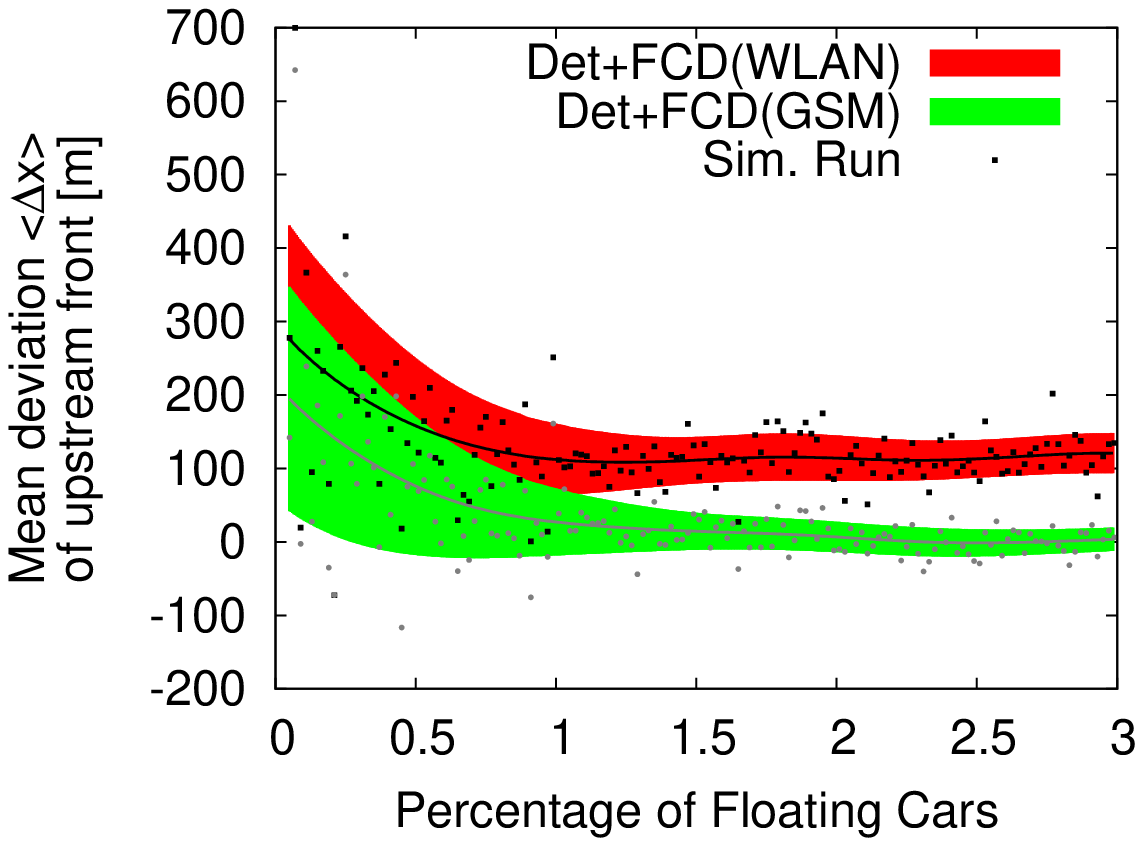}
\includegraphics[width=0.5\textwidth]{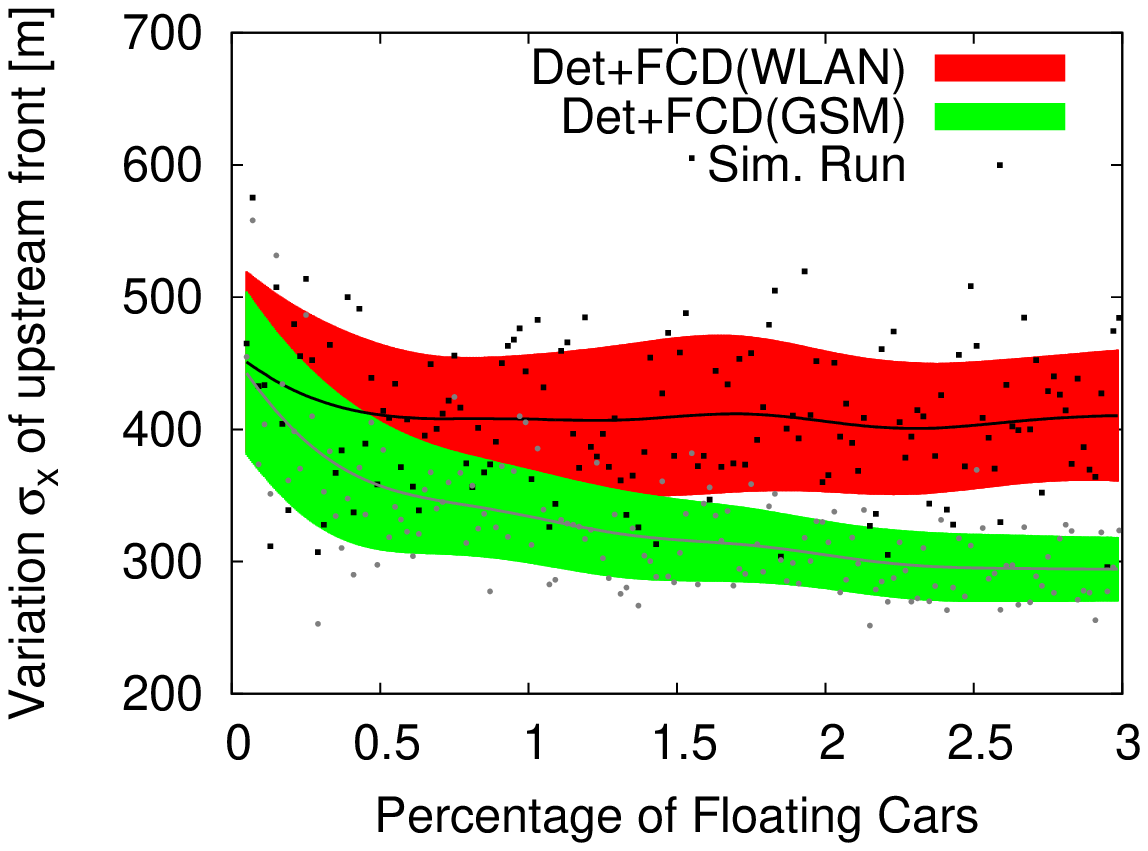} 
\end{tabular}

 \caption{\label{fig:resultsCombi}Mean error and variation of the estimated upstream jam front position for the continuity-equation based approach using flow data from stationary detectors in combination with floating car data.}

\end{figure}

Finally, we assess the prediction accuracy by means of the measure~\eqref{eq:deviation_fronts} in Fig.~\ref{fig:resultsCombi}. Compared to the  results shown in Fig.~\ref{fig:resultsFC}, the estimation model based on the continuity equation is more robust than the prediction based solely on floating cars. Note that this robustness is a prerequisite in an online application. Already for very low equipment percentages of the order of \unit[0.1]{\%},  mean deviations are about \unit[200]{m} with standard deviations of \unit[400]{m} for both FC communication modes.

\section{\label{sec:diss}Discussion and Conclusions}
%
Car-to-infrastructure communication offers interesting potentials for driver information services which are more detailed and more up-to-date than today's TMC messages and/or digital maps. For instance, infrastructure-related messages, e.g., about a temporary workzone, can be transmitted by a road-side unit (RSU) via local wireless communication devices. However, reliable information about the \emph{dynamic} traffic situation requires up-to-date traffic data and robust prediction models. In this paper, we have considered a vehicle-based approach to collect traffic data and use the data to estimate the upstream and downstream fronts of a traffic jam. A small percentage of probe vehicles records their trajectories and speeds to transmit the data to a RSU. We have shown that the delays resulting from a local communication to a single RSU positioned downstream the bottleneck will prevent a reliable jam front estimation based on floating-car data only. Obviously, the delay time  can be reduced by using more road-side units along the roadway so that probe vehicles can transmit their data in shorter intervals. 

Furthermore, we considered an alternative prediction approach using floating car data in combination with flow data collected by stationary detectors, which allows to apply the propagation equation of shock waves. Since probe vehicle data can be used for an automatic correction of the estimates and for an online calibration of the model parameters, this is a robust and promising concept which is effective already for equipment percentages of probe vehicles of a few per mill. Moreover, it is suited for a real-time application in practice.

Travel times and time losses due to congestion are also relevant dynamic quantities for local driver information services. Travel times can  directly be derived from the probe vehicle data. However, this estimation does not allow to \emph{predict} travel times which becomes relevant when no probe vehicle passes the RSU for a while. Alternatively, travel times can be calculated in a more robust way from cumulative vehicles counts. Again, probe vehicles can be used to reset the accumulated counting errors.

\paragraph{Acknowledgement:}
The authors would like to thank Holger~Poppe and Florian~Kranke for the excellent
collaboration and the Volkswagen AG for financial support within the
BMWi project AKTIV (`Adaptive and Cooperative Technologies for Intelligent Traffic'). 


\end{document}